# Characterization of the Shell Structure in Coupled Quantum Dots through Resonant Optical Probing

Mauricio Garrido[1], Kushal C. Wijesundara[1], Swati Ramanathan[1], E. A. Stinaff[1],
M. Scheibner[2], A. S. Bracker[2], D. Gammon[2]
[1]Department of Physics and Astronomy, and Nanoscale and Quantum Phenomena Institute, Ohio University, Athens, Ohio 45701-2979, USA
[2]Naval Research Laboratory, Washington, DC 20375, USA

## ABSTRACT

Excited states in single quantum dots (QDs) have been shown to be useful for spin state initialization and manipulation. For scalable quantum information processing it is necessary to have multiple spins interacting. Therefore, we present initial results from photoluminescence excitation studies of excited states in coupled quantum dots (CQDs). Due to the rich set of possible excitation and recombination possibilities, a technique for visualizing photoluminescence excitation in coupled quantum dots is discussed, by which both the interaction between the dots and the type of absorption and emission that generated the photoluminescence is easily and clearly revealed. As an example, this technique is applied to characterize the shell structure of the hole in the top dot and the results are compared with those using Level Anti-Crossing Spectroscopy (LACS).

## INTRODUCTION

QDs are often modeled as 'artificial atoms' [1,2] with a ladder of discrete energy levels arising from the 3-dimensional spatial confinement of the charge carriers. In this analogy with real atoms, the different energy levels in QDs have been equated to the shell structure in atoms including ground (s shell) and excited states (p and d shells). In CQDs the situation becomes even more intriguing. Not only are there excited states within each individual dot but also the possibility of interdot excitations and recombinations [6]. Being able to individually recognize and address the excited states in the CQDs is of great interest, and may provide a means to manipulate spin states necessary for many quantum computation schemes[3,4].

A common technique for probing the excited states of QDs is photoluminescence excitation (PLE). Traditionally, this involves identifying a peak in the photoluminescence (PL) of the QD coming from a quantum state of interest (e.g. the neutral exciton) and monitoring that specific energy while varying the energy of the laser exciting the QD. Because of the discrete nature of the states in the QD, absorption, relaxation, and emission of the laser only occurs at specific wavelengths corresponding to the excited states. This results in a series of peaks in the PLE spectrum that detail the shell structure of the quantum state of interest (Fig 1) [5].

For coupled quantum dots (CQDs), however, we need to also consider the interaction between the QDs. This interaction results in molecular states which can be altered by an applied bias voltage when the CQDs are embedded in a Schottky diode structure [6,7,8]. The dynamics of CQDs turn out to be extremely rich, with bias-dependent states and charge carriers tunneling between the dots, that PLE experiments as described above become very limited in describing the behavior of a single quantum state at different excitation energies. For example, in Figure 1 it would be difficult to differentiate between an intradot and interdot absorption. Therefore, we discuss a technique for visualizing the data that incorporates the relevant information of bias, PL intensity, PL energy and laser energy. The procedure is to create a frame, for each laser energy, that contains a series of spectra as a function of applied bias. This results in a sequence or movie which presents this multi-dimensional information in a clear and simple way, allowing for straightforward recognition of important and interesting effects. As an example, by using this technique, the shell structure hole state of the top dot is clearly indentified and characterized with the aid of the bottom dot using resonant optical probing. Comparison with Level Anti-Crossing Spectroscopy (LACS) [9] is done in order to confirm the results obtained through PLE.

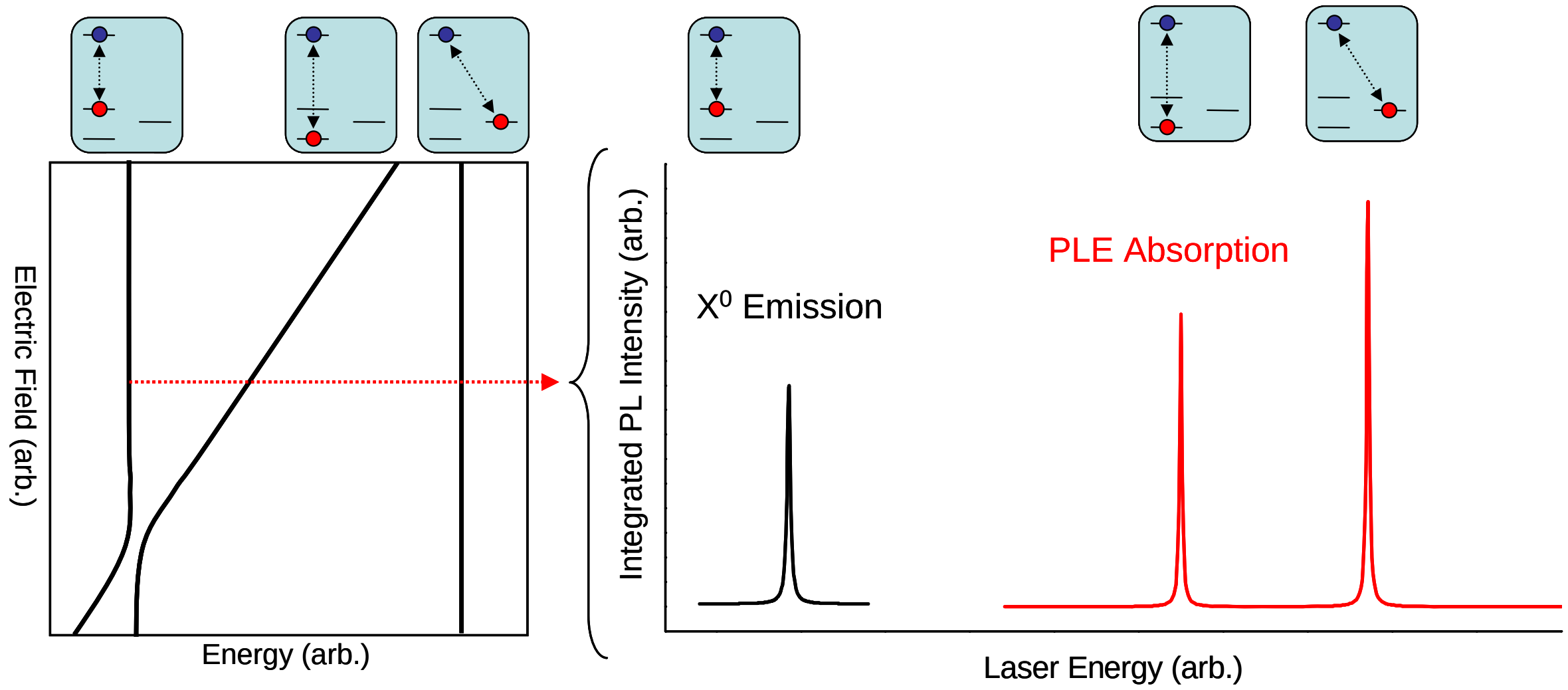

**Figure 1** Typical PLE. Left: Schematic biasmap depicting PL from a neutral exciton. Right: Representation of a PL spectra (black spectra) at a single bias (electric field), a specific PL energy is selected (in this case, that of the neutral exciton). The intensity of this line is then monitored as a function of laser energy (red spectra). Blue diagrams on top are schematics of the state of the CQD system. Blue points represent electrons, red points represent holes. Dashes on the left side of these diagrams represent the bottom dot's energy levels, while dashes on the right side represent the top dot's. In such an experiment it would be difficult to discern whether the peaks came from the recombination of direct or indirect states.

## EXPERIMENT

The systems under study are Stranski-Krastanov self-assembled GaAs/InAs coupled quantum dots embedded in an n-doped Schottky diode structure grown by MBE at the Naval Research Laboratory [10]. In order to detect single CQD spectra and apply an electric field, a semitransparent Ti top contact is deposited on the sample along with an e-beam patterned aluminum shadow mask with apertures of the order of 0.2 - 1.0 μm. The sample was cooled down to ~15K using a closed-cycle cryostat with a low vibration sample mount, keeping vibrations below 5nm. Resonant excitation was provided by a continuously tunable Ti-Sapphire CW laser. Detection was done using a triple-grating spectrometer with a maximal resolution of 0.006nm at 600nm and a liquid-nitrogen cooled, coupled charge device.

The CQD PLE data were visualized by plotting the bias, PL energy and PL intensity as the x-, y- and z-axis, respectively, of a color coded contour map, referred to as a biasmap (Figure 2). Individual biasmaps taken at different laser energies are then stacked together and viewed one after another, like frames in a movie. To compare the patterns observed in these PLE movies with measured PL emission, a static biasmap with a wider range of PL energies taken with a relatively high energy laser excitation (eg exciting above the wetting layer) is included in the PLE movie along with a moving marker (red line in Figure 2) indicating where the laser energy is for each PLE frame.

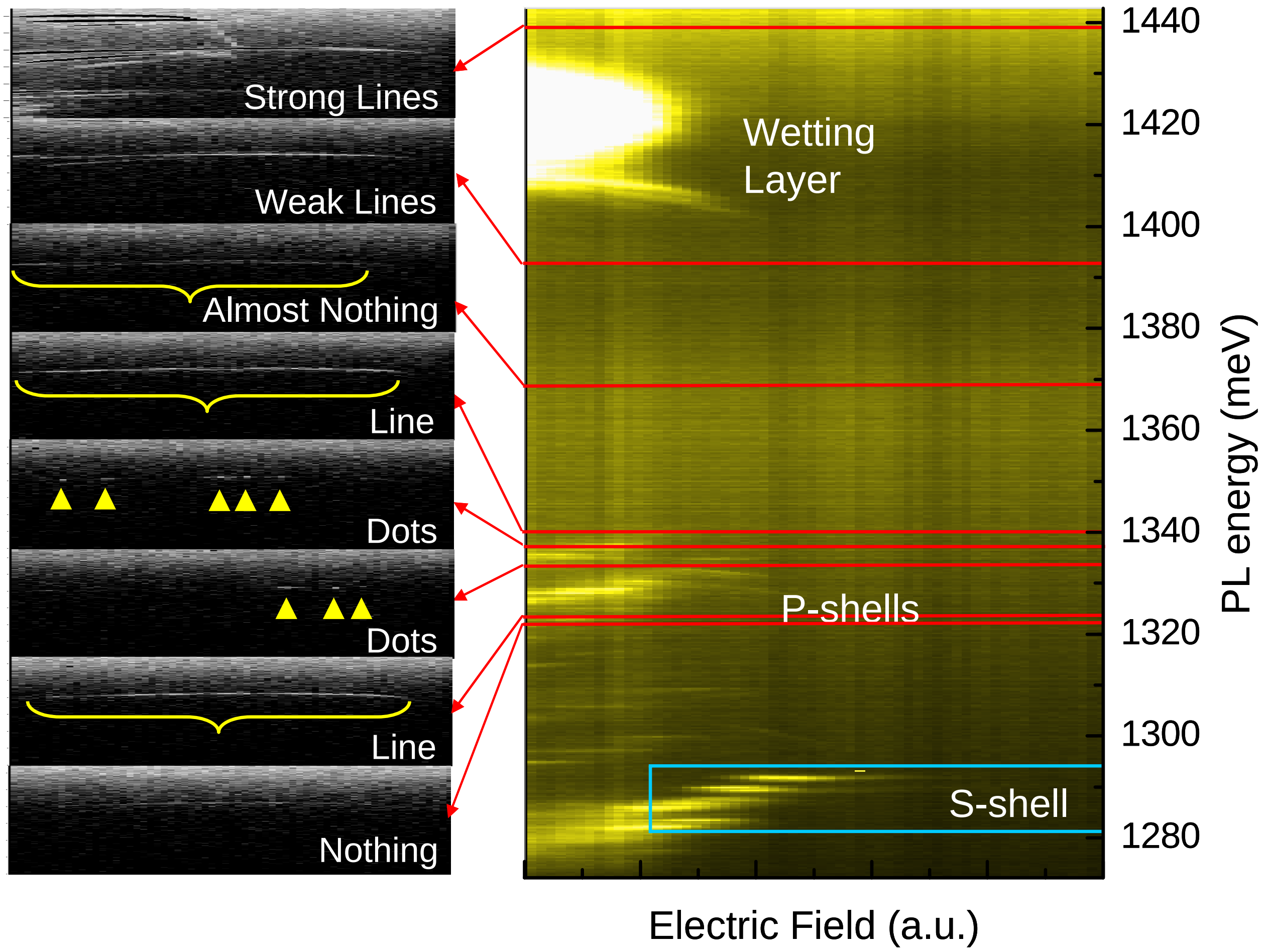

**Figure 2** Selected frames from a PLE movie. Left biasmaps show some of the observed PL signatures from the ground state of the neutral exciton as laser energy is tuned. The right biasmap is static and taken at a wider range energy range at lower resolution. The blue square indicates the energy range of left biasmaps. The red line is a the marker to show where the laser energy is for the indicated biasmap on the left.

This technique for visualizing PLE has allowed for the relatively easy identification of various PLE signatures in CQD that would otherwise have been difficult to interpret. For example, determining whether a given absorption (or emission) was due to a direct or indirect exciton would be extremely difficult (Figure 1).

## RESULTS AND DISCUSSION

As with all PLE experiments, to interpret the features in such a PLE movie we must consider the absorption, relaxation, and emission. The type of emission, direct or indirect, is revealed by the electric field dependence of the PL of the monitored ground state. The type of absorption is revealed by the way the monitored PL lines change with laser energy and electric field. Relaxation to the final states can involve both tunneling and phonon mediated relaxation.

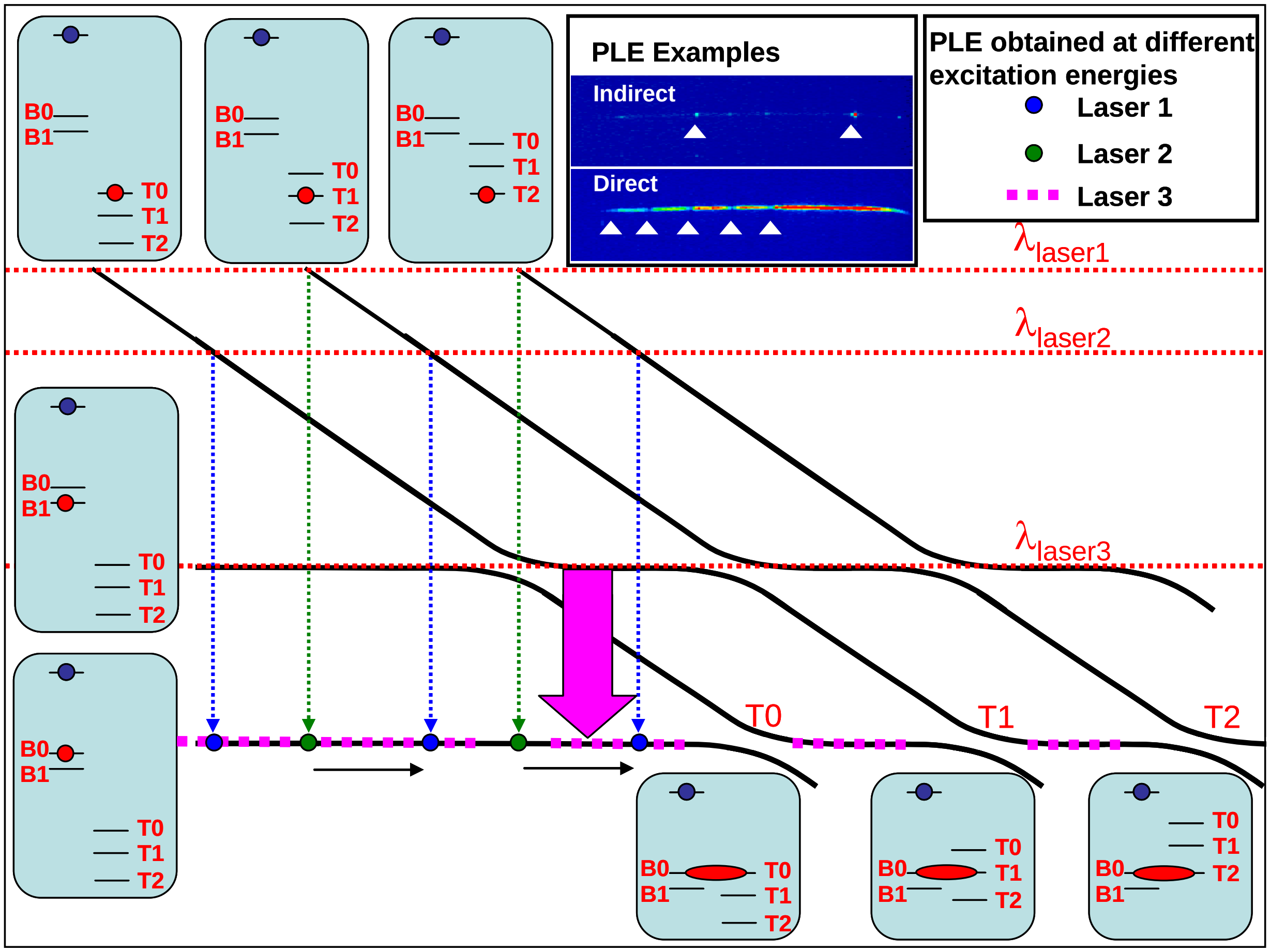

**Figure 3** Schematic representation of the processes leading to some of the PLE features in a CQD. The way the laser intersects the PL from the absorption state determines whether the PLE will be a continuous line (when exciting into a straight direct line) or a series of points (when exciting into an indirect line). The type of emission determines whether the continuous line or series of points of the PLE will be monitored along a straight line (direct recombination) or a diagonal line (indirect recombination). To show this, the above schematic depicts $\lambda_{laser1}$ and $\lambda_{laser2}$ exciting into small sections of indirect states that then tunnel, relax and recombine in the direct ground state, creating a series of points that move as the laser energy is swept from $\lambda_{laser1}$ and $\lambda_{laser2}$. On the other hand, $\lambda_{laser3}$ excites almost entirely into a direct state that then relaxes and recombines in the direct ground state, creating a mostly continuous line, with the exception maybe of a few gaps where the anticrossings are located in the PL from the absorbing state. Examples of direct and indirect types of PLE data are shown in the inset above. Arrows draw attention to series of points in indirect absorption and gaps in a line in direct absorption.

Direct exciton states, when a photon is absorbed within a single dot only, will shift only slightly due to the quantum confined Stark effect [11], therefore, direct absorption into a single dot results in a mostly bias-independent state, which in the biasmap appears as a nearly straight (single PL energy) bias-independent PL line. In certain circumstances it is possible to observe corresponding emission in a non-resonant, broad-range biasmap which may therefore be used to differentiate PL emission resulting from specific dots. Thus, we find that the PLE sequence displays emission along the entire ground state (independent of bias for direct emission, and dependent of bias for indirect emission) for a small range of laser energies. Since the state is discrete, the excitation only happens when the laser energy matches the state's energy, resulting in an extended PL line in a single frame of the movie, when the laser line is resonant with the intradot excited state. We interpret the PL observed in the inset of Figure 4 at $E_{Laser}$ = 1316.01 meV as such an intradot excitation.

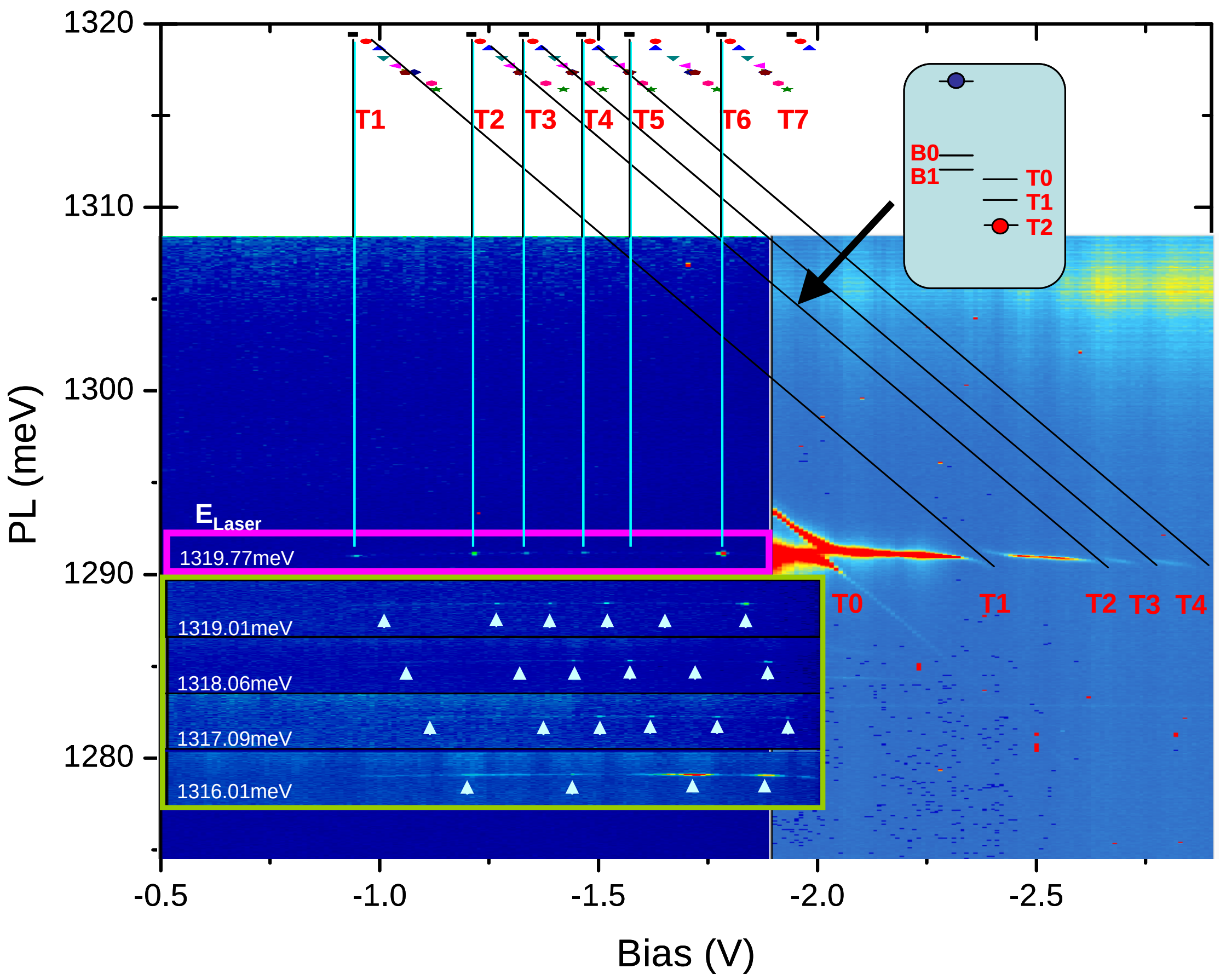

**Figure 4** PLE and LACS used to probe the shell structure of the top dot's hole. Inset is a collection of PLE frames (all centered at 1291meV) that show a series of points along the neutral exciton's ground state energy (encircled by the magenta box) moving in tandem as the laser energy is changed. Arrow markers have been placed to better show the succession of points. When these points are plotted at the laser energy that created them, a series of lines is built up, which match the diagonal PL lines from the indirect excited states seen using LACS. It is interesting to note that the PL biasmap taken at Elaser = 1316.01 meV is indicative of a direct recombination.

Indirect transitions result in highly bias-dependent states since the energy levels of the top dot shift with respect to the bottom dot energy levels as the bias is changed. In a single PL biasmap with non resonant excitation, this appears as a diagonal bias-dependent emission line, as represented in Figure 3. In a PLE experiment, the laser can excite into this level only when its energy matches that of the indirect state at a specific bias. Therefore, after tunneling and relaxation, this is seen as a single PL point in the biasmap at the ground state emission energy. As the laser energy changes, the bias at which it can excite into the indirect state also changes, shifting the electric field at which the PL point is observed. Thus, a series of points is built up frame-by-frame, making it appear as a moving point along the bias-independent line of the direct, ground state emission (Figure 4).

We apply this knowledge now to characterizing the shell structure of the top dot by creating a PLE movie of the energy range near the ground state of the neutral exciton in the bottom dot. Similar to the signature discussed in the previous section, a couple of points appear along the neutral exciton's straight bias-independent line, all moving in tandem as the laser energy is changed (See inset Figure 4). Therefore, it is clear that these come from an indirect absorption and a direct recombination. Also, from this last statement, it can be deduced that there must have been tunneling at some point in the dynamics of the created indirect neutral exciton to have ended up as a direct neutral exciton.

The points are inferred to have come from the indirect absorption into excited states of the indirect neutral exciton. Therefore, the separation between these points should reveal the shell structure of the hole in the top dot. To verify this assumption, LACS was performed on the CQD, where the electric fields at which the top dots excited states become resonant with the bottom dot are revealed as a series of anticrossings in the exciton PL as the electric field is increases [9]. This not only shows the separation between the different excited states, but also their position in the biasmap. Graphing the points where the laser created the PLE points, produced a series of lines that were then interpolated and seen to match the anti-crossings associated with the excited states of the top dot, confirming the assumption of the PLE signal resulting from the excited states of the hole in the top dot (Figure 4). One advantage of the PLE experiment is that, although the ground state PL where the level anti-crossings are observed becomes much weaker as the electric field increases making their recognition more difficult, the resonant excitation into the excited states results in relatively high signal intensities even for high lying excited states.

## CONCLUSIONS

We have optically probed the shell structure of the top dot by resonantly creating an indirect neutral exciton and monitoring the direct neutral exciton. This was achieved using the understanding obtained from the new technique to visualize PLE. The identification of these excited states was given by relating the results obtained using PLE with those using LACS.

## ACKNOWLEDGMENTS

The authors would like to thank CMSS for their support.